\documentclass[sigconf]{acmart}
\AtBeginDocument{%
  }

\setcopyright{acmlicensed}
\copyrightyear{2018}
\acmYear{2025}
\acmDOI{XXXXXXX.XXXXXXX}
\acmConference[ACM-BCB '25]{16th ACM Conference on Bioinformatics, Computational Biology, and Health Informatics}{October 11--15, 2025}{Philadelphia, PA}

\usepackage[table]{xcolor}
\definecolor{mincolor}{RGB}{255,255,255} 
\definecolor{maxcolor}{RGB}{178,34,34}   

\usepackage{booktabs}
\usepackage{csvsimple}
\usepackage{pgf}
\usepackage{pgfmath}
\usepackage{tabularx}
\usepackage[font=small]{caption}
\usepackage{balance}

\copyrightyear{2025}
\acmYear{2025}
\setcopyright{cc}
\setcctype{by}
\acmConference[BCB Companion '25]{16th ACM International Conference on
Bioinformatics, Computational Biology and Health Informatics}{October 11--15,
2025}{Philadelphia, PA, USA}
\acmBooktitle{16th ACM International Conference on Bioinformatics,
Computational Biology and Health Informatics (BCB Companion '25), October
11--15, 2025, Philadelphia, PA, USA}
\acmDOI{10.1145/3768322.3769032}
\acmISBN{979-8-4007-2222-6/2025/10}

\begin{document}
\title{PRIMRose: Insights into the Per-Residue Energy Metrics of Proteins with Double InDel Mutations using Deep Learning}

\author{Stella Brown}
\affiliation{%
  \institution{Western Washington University}
  \city{Bellingham}
  \country{USA}}
\email{browns63@wwu.edu}

\author{Nicolas Preisig}
\affiliation{%
  \institution{Western Washington University}
  \city{Bellingham}
  \country{USA}}
\email{preisin@wwu.edu}

\author{Autumn Davis}
\affiliation{%
  \institution{Western Washington University}
  \city{Bellingham}
  \country{USA}}
\email{davisd33@wwu.edu}

\author{Brian Hutchinson}
\affiliation{%
  \institution{Western Washington University}
  \city{Bellingham}
  \country{USA}}
\email{Brian.Hutchinson@wwu.edu}

\author{Filip Jagodzinski}
\affiliation{%
  \institution{Western Washington University}
  \city{Bellingham}
  \country{USA}}
\email{filip.jagodzinski@wwu.edu}

\begin{abstract}
  Understanding how protein mutations affect protein structure is essential for advancements in computational biology and bioinformatics. We introduce PRIMRose, a novel approach that predicts energy values for each residue given a mutated protein sequence. Unlike previous models that assess global energy shifts, our method analyzes the localized energetic impact of double amino acid insertions or deletions (InDels) at the individual residue level, enabling residue-specific insights into structural and functional disruption. We implement a Convolutional Neural Network architecture to predict the energy changes of each residue in a protein mutation. We train our model on datasets constructed from nine proteins, grouped into three categories: one set with exhaustive double InDel mutations, another with approximately 145k randomly sampled double InDel mutations, and a third with approximately 80k randomly sampled double InDel mutations. Our model achieves high predictive accuracy across a range of energy metrics as calculated by the Rosetta molecular modeling suite and reveals localized patterns that influence model performance, such as solvent accessibility and secondary structure context. This per-residue analysis offers new insights into the mutational tolerance of specific regions within proteins and provides higher interpretable and biologically meaningful predictions of InDels' effects. 
\end{abstract}



\received{}
\received[revised]{}
\received[accepted]{}

\maketitle

\section{Introduction}
Wet lab studies are a fundamental step in the study of the effects of protein mutations; however, the process of performing these experiments requires a substantial amount of time, energy, and lab resources to carry out even a single amino acid mutation.
Experimentally testing the effects of every possible protein mutation is infeasible due to the immense time, cost, and labor required. Without computational guidance, most laboratory efforts risk experimenting with mutations that yield little to no insight. Computational methods - such as the Rosetta software suite - simulate structural changes and assess the energetic changes in mutations. Rosetta calculates a set of energy terms, ``Rosetta scores'', to describe properties such as stability, solvation, and intra- or inter-residue interactions to provide a standardized guideline for predicting mutation effects \cite{Rosetta}. 

This project utilizes a previously established computation pipeline \cite{Turcan2022} that leverages Rosetta's ability to generate \textit{in silico} double InDel mutations across nine protein structures in PDB format. For larger proteins, exhaustively generating all possible double InDel variants - possibly surmounting two million mutants - can require over 3,000 CPU-days. Even when distributed across hundreds of CPUs, these computations can still take several weeks to complete, as noted in Coffland {\it et al.} (2023). These calculations do not significantly eliminate the resources needed compared to wet lab studies. 

In this work, we propose a novel approach to predict the effects of mutations on a per-residue level for a given protein. Given a FASTA sequence of amino acid identities for a protein, our model generates these Rosetta scores for every residue location. We train a convolutional neural network called PRIMRose (Per-Residue InDel Modeling with Rosetta) to capture underlying contextual patterns in proteins, including interactions between neighboring residues and variations in solvent-accessible surface area (SASA) measurements. Expanding this scope allows us to analyze the structural impacts of insertions that reside in a protein's secondary structure. 

\section{Related Work}
Research efforts have heavily relied on wet-lab studies and began incorporating computational modeling tools in the 1970s to aid in these pursuits \cite{Drake1976, Masso2008, Moreira2007}; however, research focusing on the effects of InDels is less thoroughly explored. 

Recent computational efforts have advanced the field significantly. For example, Jilani {\it et al.} explored the structural impacts of InDels on disease phenotypes through modeling techniques \cite{Jilani2022}. Complementing this, Topolska {\it et al.} conducted deep InDel mutagenesis and developed INDELi, a model that integrates substitution and structural context to predict InDel effects on stability and functionality \cite{Topolska2025}. This work revealed the complexity and variability of InDel tolerance across domains. Miton and Tokuriki emphasized the evolutionary potential of InDels, highlighting that insertions or deletions can trigger large structural changes in proteins that are often impossible to achieve through amino acid substitution alone \cite{Miton2022}. 

Several machine learning (ML) methods have also been developed to predict the impacts of mutations. Diaz {\it et al.} reviewed the staggering progress of ML in modeling mutational effects, particularly using convolution, recurrent, and transformer-based neural networks to predict protein fitness, stability, and optimal design variants \cite{Diaz2022}. Similarly, Zhou {\it et al.} introduced DDMut, a deep learning model using Siamese networks and graph-based encodings, which outperforms traditional tools across mutation types. It is important to acknowledge the limitations in these ML models; Fang identified how overfitting and poor generalizability highlight a need for more robust training datasets and validation through reverse mutation strategies \cite{Fang2019}. 

Older statistical and hybrid approaches continue to remain relevant. Earlier approaches examined correlations in multiple sequence alignments of amino acids to predict contact maps for protein families \cite{Gobel1994}. These diagrams display regions of a protein that interact spatially with each other. Pandurangan and Blundell's mCSM and SDM frameworks, and the ensemble DUET model, have been highly influential in predicting mutation impacts using both statistical potentials and ML features, particularly for drug resistance and stability predictions \cite{Pandurangan2019}. In addition, GEMME - a global epistatic model - uses evolutionary trace information to rapidly and accurately infer mutational effects by considering phylogenetic conservation and sequence-wide epistasis \cite{Laine2019}. Other tools such as SIMPROT simulate sequence evolution under InDel and substitution pressures \cite{Pang2005}. Despite these advances, van der Flier {\it et al.} note that predicting mutational effects remains an open challenge that requires more integrative and novel strategies \cite{Flier2023}.

In contrast to these prior approaches and models, which generally predict global mutational effects and residue-level summaries across entire proteins, our new per-residue model offers position-specific predictions of mutational energy changes. Unlike DDMut or GEMME, which focus on mutation type or evolutionary coupling, our model evaluates structural and energetic changes at every residue position using localized Rosetta-derived energy terms. By calculating energy changes at each individual residue, our model captures how different parts of a protein respond differently to mutations. This spatial heterogeneity is especially important for InDels since their effects can vary greatly depending on where they occur within the protein structure. This enables a higher-resolution mapping of mutational landscapes, revealing specific structural regions that are tolerant or vulnerable to mutations - insights that are often missed in models trained on whole-protein summary statistics. 

The RoseNet model developed by Coffland {\it et al.} \cite{rosenet, rosenet_extension} largely serves as the inspiration for our model's task. RoseNet and PRIMRose differ significantly in architecture, but both are trained to predict Rosetta scores of double InDel mutants, with the major difference being that RoseNet makes a single whole-protein prediction per score, whereas PRIMRose predicts each score per residue.

Like our approach, RoseNet consists of an independent model trained on each protein, each of which accepts the parameters of a double InDel mutation as input in the space $ \mathbb{Z}^4 $, representing the first and second insertion positions within the wildtype, and the first and second amino acid identity, respectively, that define the mutation. Unlike PRIMRose, a RoseNet model predicts a vector of Rosetta scores for each mutant, representing the \textit{summation over residue} of each score value \cite{rosenet}.

The RoseNet architecture consists of an embedding layer that embeds all four inputs into vector representations, followed by one or more "RoseNet blocks", which consist of multiple fully-connected linear layer, batch norm, ReLU activation sequences, using residual connections \cite{rosenet}.

Coffland {\it et al.} trained, tested, and published result for RoseNet on exhaustive double InDel mutation datasets for three proteins: 1crn, 1csp, and 1hhp \cite{rosenet}. The authors later published an extension in which they trained, tested, and published results on the same dataset in addition to ~145k randomly selected double InDel mutations each for three additional proteins: 2ckx, 1c44, and 5cvz \cite{rosenet_extension}. In each protein they achieved statistically significant results, with Pearson correlations between predicted and actual scores frequently within the range of 0.6-0.7.
\section{Methods}
\subsection{Data Generation}
We train our model on mutations generated for nine protein structures from the Protein Data Bank (PDB): 1crn, 1csp, and 1hhp— originally used in the first RoseNet publication \cite{rosenet} — and six additional proteins: 2ckx, 1c44, 5cvz, 1wm2, 1dc9, and 7dzu. We follow the approach of RoseNet by using exhaustive datasets containing all possible double InDel mutations for 1crn, 1csp, and 1hhp, and likewise the approach of RoseNet Extension \cite{rosenet_extension} by generating approximately 145k random mutants for 2ckx, 1c44, and 5cvz. For the remaining proteins, 1wm2, 1dc9, and 7dzu, we generate approximately 80k mutations each. We find that these models are capable of learning from only fraction of the exhaustive sets of all mutations used by the original RoseNet, demonstrating that PRIMRose, like RoseNet, can be replicated on new proteins with less computation. Individual sequence lengths and mutation counts are included in Table~\ref{tab:num-mutations-exact}.
\begin{table}
    \centering
    \small
    \resizebox{\linewidth}{!}{%
    \begin{tabular}{|c|ccc|ccc|ccc|}
        \hline
        \textbf{} & \multicolumn{3}{c|}{Exhaustive} & \multicolumn{3}{c|}{$\sim$145k Mutations} & \multicolumn{3}{c|}{$\sim$80k Mutations} \\
        Protein & 1crn & 1csp & 1hhp & 2ckx & 1c44 & 5cvz & 1wm2 & 1dc9 & 7dzu \\
        \hline
        Protein Length & 46 & 67 & 99 & 83 & 123 & 141 & 78 & 131 & 268 \\
        Num Mutations & 447,253 & 939,756 & 1,529,491 & 144,755 & 142,611 & 147,209 & 75,033 & 78,729 & 89,991 \\
  \hline
    \end{tabular}
    }
    \caption{Sequence lengths of each protein before double insertions and number of mutations generated for each protein.}
    \label{tab:num-mutations-exact}
\end{table}
\subsection{Data Splits}
Before creating traditional train/validation/test data splits, we hold out two additional test sets ($\text{Test}_{\text{Pos}}$ and $\text{Test}_{\text{AA}}$) to determine the ability of our model to generalize to unseen insertion position pairs and amino acid pairs, respectively. We hold out 5\% of our original data for $\text{Test}_{\text{Pos}}$ and 5\% for $\text{Test}_{\text{AA}}$. We then split the remaining data randomly at 80\%-10\%-10\% for our training, validation, and $\text{Test}_{\text{Rand}}$ sets, respectively. $\text{Test}_{\text{Pos}}$ contains pairs of insertion positions that are not present in the training, validation, and serves as a way to view how the model handles inputs that are substantially different from the training data. $\text{Test}_{\text{AA}}$ contains amino acid combinations that are not present in training, validation, or $\text{Test}_{\text{Rand}}$ datasets and shows how the model responds to amino acid pairs it was not trained on. Both of the datasets provide insight into the robustness of PRIMRose's understanding of the effects of mutations.
$\text{Test}_{\text{Rand}}$ acts as a traditional test dataset with data that remains unseen during training but contains amino acid pairs and insertion position pairs that are present in the training and validation data. These three test sets allow us to measure the model's ability to generalize to unseen data, novel pairs of insertion positions, and novel pairs of inserted amino acids.
\subsection{Rosetta Score Selection and Preprocessing}
Each model is trained to predict 14 of the 20 energy metrics calculated by Rosetta. We observed that both dslf\_fa13 and yhh\_planarity remained effectively constant across all mutations for each protein, suggesting that these terms do not meaningfully differentiate mutational effects in our dataset. We chose to omit these terms as they would offer negligible signal for model learning and could bias training toward irrelevant patterns. Additionally, all hydrogen bond-related scores (hbond\_sr\_bb, hbond\_lr\_bb, hbond\_bb\_sc, and hbond\_sc) are excluded from training and evaluation. These terms are inherently pairwise interaction energies that depend on the geometric relationship between two residues or atoms, rather than on properties localized to a single residue. When generating all mutations, we did not run the computation pipeline \cite{Turcan2022} with the decompose\_bb\_hb\_into\_pair\_energies flag set to true \cite{RosettaScripts}; therefore, Rosetta frequently outputs these scores as zeros for most mutations, indicating that a mutation often did not participate in a hydrogen bond within the localized residue window. As a result, these scores provide sparse signals that are poorly suited to our model's per-residue, position-independent framework and would likely introduce noise.
Table~\ref{tab:rosetta_metrics} summarizes the 14 Rosetta scores we used in our evaluations.

\begin{table}
    \centering
    \small
    \begin{tabular}{@{}l p{5cm}@{}}
        \toprule
        \textbf{Rosetta score/metric} & \textbf{Meaning} \\
        \midrule
        \texttt{fa\_atr} & Lennard-Jones attractive force between atoms in different residues \\
        \texttt{fa\_rep} & Lennard-Jones repulsive force between atoms in different residues \\
        \texttt{fa\_sol} & Lazaridis-Karplus solvation energy \\
        \texttt{fa\_intra\_sol} & Intra-residue Lazaridis-Karplus solvation energy \\
        \texttt{lk\_ball\_wtd} & Asymmetric solvation energy \\
        \texttt{fa\_intra\_rep} & Lennard-Jones repulsion between atoms in the same residue \\
        \texttt{fa\_elec} & Coulombic electrostatic potential with distance-dependent dielectric \\
        \texttt{pro\_close} & Proline ring closure energy and energy of $\psi$ angle of preceding residue \\
        \texttt{rama\_prepro} & Ramachandran preferences \\
        \texttt{omega} & Omega dihedral in the backbone \\
        \texttt{p\_aa\_pp} & Probability of amino acid, given torsion values for $\phi$ and $\psi$ \\
        \texttt{fa\_dun} & Internal energy of sidechain rotamers \\
        \texttt{ref} & Reference energy for each amino acid \\
        \texttt{total} & The total weighted score for the structure \\
        \bottomrule
    \end{tabular}
    \vspace{1em}
    \caption{Energy metrics output by Rosetta.}
    \label{tab:rosetta_metrics}
\end{table}
\subsection{Model Inputs and Architecture}
This per-residue level model must predict Rosetta scores for every amino acid in the protein sequence; therefore, we choose to adopt a Convolutional Neural Network (CNN) architecture with residual connections. This fully-convolutional architecture is better suited than a fully-connected deep neural network (e.g., as used in RoseNet) for making per-residue predictions over a sequence as it can be applied to arbitrarily sized sequences without changing the model and its output scales with the size of the input. A convolutional layer consists of a set of filters that are applied at different positions of the input sequence with the goal of extracting important features of the input. For our purposes, the convolutional layers we employ are one dimensional and have a kernel width of 3, stride of 1, and padding of 1; as a result, each convolutional layer preserves the spatial dimensions. Our architecture is made up of 30 residual blocks that keep a channel dimension of 64 followed by a final convolutional output layer that brings the channel dimension down to 14, the number of scores we are predicting. Each of the residual blocks is made up of two sequences, each consisting of a convolutional layer, a ReLU activation, and a BatchNorm layer. The architecture of one block is expressed in the equation below. The input to the model is the sequence of amino acids comprising the mutant which are then each embedded as vectors of size 64. The final input to the model is therefore a $\text{[batch size} \times \text{sequence length} \times 64]$ tensor. To improve training, the target scores are normalized using z-score normalization based on the means of the training data.
\begin{equation}
  \begin{aligned}
    h_1 &= ReLU(BatchNorm_1(Conv_1(x))) \\
    h_2 &= ReLU(BatchNorm_2(Conv_2(y_1))) \\
    y &= h_2 + x
    \end{aligned}
\end{equation}

\subsection{Training and Tuning Procedure}
We train our model using the Adam optimizer \cite{ADAM} with an initial learning rate of 0.0001. The loss function we use is Huber loss, chosen to reduce the influence of outliers in the target scores. Training is conducted with a batch size of 256 and a performance learning rate scheduler with a factor of 0.1 and a patience of 10. The batch size, learning rate, and number of residual blocks are tuned to optimize 1crn's validation set performance using Bayesian hyperparameter optimization of the hyperparameter space via WandB \cite{wandb}. For simplicity, we use the same hyperparameters across all proteins, finding that they work well across this range. Performance may be further improved if independent hyperparameter searches were conducted for each protein. 

All training is performed on systems with an NVIDIA RTX 4090 GPU and 32 GB of main memory. This is the minimum amount of memory required to fit each dataset entirely into RAM due to 1hhp's size and exhaustive number of mutations -- ie., allowing eager loading, which speeds up training significantly.

Each protein is trained for 100 epochs while checkpointing the best model parameters determined by highest Pearson correlation coefficients across the residues on the validation set; models for each protein converge by epoch 40. The model does not show signs of overfitting.

\section{Results}
Model performance is assessed by calculating the Pearson correlation coefficient between predicted and ground-truth Rosetta scores. Let $\mathcal{P}$ denote our set of nine proteins, and $p \in \mathcal{P}$ denote a specific protein. Let $\mathcal{S}$ denote the set of 14 Rosetta scores, and let $s \in \mathcal{S}$ denote a specific score. Let $L_p$ denote the length (in residues) of protein $p$, and let $N_p$ denote the number of mutants in the given set for the protein (e.g., $\text{Test}_{\text{Pos}}$).
Finally, let $\hat{y}_{psmi}$ denote PRIMRose's prediction for Rosetta score $s$ of protein $m$ at position $i$ of the $m$th mutation of the protein. Let $y_{psmi}$ denote the corresponding ground truth value from Rosetta. Then for each protein $p$ and each of the 14 scores, we calculate a score-specific Pearon correlation coefficient, $r_{ps}$, as follows:
\begin{equation}
r_{ps} = \frac{\sum_{m=1}^{N_p}\sum_{i=1}^{L_p}{(\hat{y}_{psmi}-\hat{\mu}_{ps})(y_{psmi}-\mu_{ps})}} 
{
\sqrt{\sum_{m=1}^{N_p}\sum_{i=1}^{L_p}(\hat{y}_{psmi}-\hat{\mu}_{ps})^2}
\sqrt{\sum_{m=1}^{N_p}\sum_{i=1}^{L_p}(y_{psmi}-\mu_{ps})^2}
}.
\end{equation}
Here $\hat{\mu}_{ps}$ denotes the mean over the $N_p L_p$ predictions for protein $p$ and score $s$, and $\mu_{ps}$ denotes the equivalent mean over the ground truth values.
Pearson correlation coefficients capture the strength and direction of linear correlation (higher is better), making them useful for evaluating prediction performance.

Table~\ref{tab:test1_pearsons} contains a color-coded heatmap that highlights performance across different proteins and scores. A clear pattern emerges with energy terms associated with local interactions, such as fa\_atr, fa\_sol, and fa\_intra\_sol, which consistently yield high correlations across nearly all proteins. Many scores exceed a Pearson of 0.9 with particularly strong correlations in proteins with longer sequences. For example, fa\_sol reaches values above 0.98 in multiple proteins including 1csp, 1hhp, and 7dzu. 

Conversely, energy terms associated with nonlocal or structurally complex interactions, such as pro\_close and total, show lower and more variable performance. These values may reflect the difficulty in modeling long-range interactions or cooperativity using individual residue features alone. 

Overall, the heatmaps reveal that our model effectively captures local energy changes with this residue resolution, with performance scaling positively with protein length. 

We see highly consistent heatmap color patterns when our model evaluates previously unseen insertion positions combinations and amino acid combinations (Tables~\ref{tab:test2_pearsons},~\ref{tab:test3_pearsons}). These patterns highlight our model's stable predictive performance across both seen and unseen mutation conditions. Notably, energy terms such as fa\_atr, fa\_sol, and fa\_intra\_sol maintain strong correlations across test sets. This consistency emphasizes our model's ability to generalize to novel mutations, suggesting that the learned representations capture fundamental features of residue-level energy changes rather than overfitting to specific mutation types or datasets. 
\begin{table}
    \centering
    \small
    \caption{Pearson correlation coefficients results for all proteins ($\text{Test}_{\text{Rand}}$).}
    \label{tab:test1_pearsons}
    \resizebox{\linewidth}{!}{%
    \begin{tabular}{|c|ccc|ccc|ccc|}
        \hline
        \textbf{Rosetta Score} & \multicolumn{3}{c|}{Exhaustive} & \multicolumn{3}{c|}{$\sim$145k Mutations} & \multicolumn{3}{c|}{$\sim$80k Mutations} \\
        & 1crn & 1csp & 1hhp & 2ckx & 1c44 & 5cvz & 1wm2 & 1dc9 & 7dzu \\
        \hline
        fa\_atr & \cellcolor[RGB]{252,254,164}\textcolor{black}{0.997} & \cellcolor[RGB]{252,254,164}\textcolor{black}{0.999} & \cellcolor[RGB]{252,254,164}\textcolor{black}{0.999} & \cellcolor[RGB]{242,228,93}\textcolor{black}{0.947} & \cellcolor[RGB]{242,228,93}\textcolor{black}{0.948} & \cellcolor[RGB]{243,226,89}\textcolor{black}{0.946} & \cellcolor[RGB]{250,191,37}\textcolor{black}{0.893} & \cellcolor[RGB]{243,222,82}\textcolor{black}{0.939} & \cellcolor[RGB]{244,247,141}\textcolor{black}{0.982} \\
        fa\_rep & \cellcolor[RGB]{237,106,35}\textcolor{black}{0.761} & \cellcolor[RGB]{244,122,22}\textcolor{black}{0.789} & \cellcolor[RGB]{231,95,44}\textcolor{black}{0.742} & \cellcolor[RGB]{88,16,109}\textcolor{white}{0.465} & \cellcolor[RGB]{85,15,109}\textcolor{white}{0.460} & \cellcolor[RGB]{26,11,64}\textcolor{white}{0.361} & \cellcolor[RGB]{0,0,3}\textcolor{white}{0.280} & \cellcolor[RGB]{8,6,31}\textcolor{white}{0.321} & \cellcolor[RGB]{80,13,108}\textcolor{white}{0.451} \\
        fa\_sol & \cellcolor[RGB]{252,254,164}\textcolor{black}{0.998} & \cellcolor[RGB]{252,254,164}\textcolor{black}{0.998} & \cellcolor[RGB]{252,254,164}\textcolor{black}{0.999} & \cellcolor[RGB]{242,228,93}\textcolor{black}{0.948} & \cellcolor[RGB]{242,230,96}\textcolor{black}{0.949} & \cellcolor[RGB]{242,228,93}\textcolor{black}{0.947} & \cellcolor[RGB]{251,179,24}\textcolor{black}{0.878} & \cellcolor[RGB]{244,219,75}\textcolor{black}{0.934} & \cellcolor[RGB]{244,247,141}\textcolor{black}{0.982} \\
        fa\_intra\_rep & \cellcolor[RGB]{246,250,149}\textcolor{black}{0.987} & \cellcolor[RGB]{250,253,160}\textcolor{black}{0.994} & \cellcolor[RGB]{250,253,160}\textcolor{black}{0.996} & \cellcolor[RGB]{249,199,47}\textcolor{black}{0.906} & \cellcolor[RGB]{246,128,18}\textcolor{black}{0.800} & \cellcolor[RGB]{251,164,10}\textcolor{black}{0.853} & \cellcolor[RGB]{248,136,12}\textcolor{black}{0.812} & \cellcolor[RGB]{248,136,12}\textcolor{black}{0.811} & \cellcolor[RGB]{249,197,44}\textcolor{black}{0.902} \\
        fa\_intra\_sol & \cellcolor[RGB]{249,252,157}\textcolor{black}{0.994} & \cellcolor[RGB]{252,254,164}\textcolor{black}{0.997} & \cellcolor[RGB]{252,254,164}\textcolor{black}{0.999} & \cellcolor[RGB]{241,237,112}\textcolor{black}{0.962} & \cellcolor[RGB]{241,233,104}\textcolor{black}{0.956} & \cellcolor[RGB]{241,240,121}\textcolor{black}{0.966} & \cellcolor[RGB]{244,219,75}\textcolor{black}{0.933} & \cellcolor[RGB]{241,235,108}\textcolor{black}{0.957} & \cellcolor[RGB]{247,251,153}\textcolor{black}{0.990} \\
        lk\_ball\_wtd & \cellcolor[RGB]{244,247,141}\textcolor{black}{0.982} & \cellcolor[RGB]{243,246,137}\textcolor{black}{0.978} & \cellcolor[RGB]{242,228,93}\textcolor{black}{0.948} & \cellcolor[RGB]{251,181,26}\textcolor{black}{0.881} & \cellcolor[RGB]{250,193,40}\textcolor{black}{0.897} & \cellcolor[RGB]{250,191,37}\textcolor{black}{0.893} & \cellcolor[RGB]{247,133,14}\textcolor{black}{0.806} & \cellcolor[RGB]{248,203,52}\textcolor{black}{0.911} & \cellcolor[RGB]{243,226,89}\textcolor{black}{0.943} \\
        fa\_elec & \cellcolor[RGB]{247,251,153}\textcolor{black}{0.989} & \cellcolor[RGB]{246,250,149}\textcolor{black}{0.987} & \cellcolor[RGB]{249,252,157}\textcolor{black}{0.993} & \cellcolor[RGB]{244,220,79}\textcolor{black}{0.936} & \cellcolor[RGB]{243,224,86}\textcolor{black}{0.941} & \cellcolor[RGB]{247,207,58}\textcolor{black}{0.918} & \cellcolor[RGB]{251,160,7}\textcolor{black}{0.849} & \cellcolor[RGB]{245,215,69}\textcolor{black}{0.927} & \cellcolor[RGB]{242,244,133}\textcolor{black}{0.976} \\
        pro\_close & \cellcolor[RGB]{147,37,103}\textcolor{white}{0.567} & \cellcolor[RGB]{134,33,106}\textcolor{white}{0.546} & \cellcolor[RGB]{190,56,82}\textcolor{white}{0.647} & \cellcolor[RGB]{62,9,102}\textcolor{white}{0.420} & \cellcolor[RGB]{64,9,102}\textcolor{white}{0.421} & \cellcolor[RGB]{26,11,64}\textcolor{white}{0.359} & \cellcolor[RGB]{64,9,102}\textcolor{white}{0.422} & \cellcolor[RGB]{134,33,106}\textcolor{white}{0.545} & \cellcolor[RGB]{120,28,109}\textcolor{white}{0.521} \\
        omega & \cellcolor[RGB]{247,207,58}\textcolor{black}{0.915} & \cellcolor[RGB]{250,145,7}\textcolor{black}{0.828} & \cellcolor[RGB]{250,253,160}\textcolor{black}{0.995} & \cellcolor[RGB]{59,9,100}\textcolor{white}{0.414} & \cellcolor[RGB]{243,222,82}\textcolor{black}{0.939} & \cellcolor[RGB]{137,34,105}\textcolor{white}{0.552} & \cellcolor[RGB]{175,49,91}\textcolor{white}{0.619} & \cellcolor[RGB]{136,33,106}\textcolor{white}{0.549} & \cellcolor[RGB]{247,251,153}\textcolor{black}{0.991} \\
        fa\_dun & \cellcolor[RGB]{250,253,160}\textcolor{black}{0.994} & \cellcolor[RGB]{250,253,160}\textcolor{black}{0.996} & \cellcolor[RGB]{252,254,164}\textcolor{black}{0.998} & \cellcolor[RGB]{241,235,108}\textcolor{black}{0.957} & \cellcolor[RGB]{242,228,93}\textcolor{black}{0.946} & \cellcolor[RGB]{243,222,82}\textcolor{black}{0.939} & \cellcolor[RGB]{243,226,89}\textcolor{black}{0.945} & \cellcolor[RGB]{244,219,75}\textcolor{black}{0.934} & \cellcolor[RGB]{245,248,145}\textcolor{black}{0.983} \\
        p\_aa\_pp & \cellcolor[RGB]{180,51,88}\textcolor{white}{0.626} & \cellcolor[RGB]{250,193,40}\textcolor{black}{0.896} & \cellcolor[RGB]{247,207,58}\textcolor{black}{0.916} & \cellcolor[RGB]{243,119,25}\textcolor{black}{0.784} & \cellcolor[RGB]{247,207,58}\textcolor{black}{0.918} & \cellcolor[RGB]{251,176,20}\textcolor{black}{0.871} & \cellcolor[RGB]{226,87,51}\textcolor{white}{0.725} & \cellcolor[RGB]{248,205,55}\textcolor{black}{0.914} & \cellcolor[RGB]{243,226,89}\textcolor{black}{0.944} \\
        ref & \cellcolor[RGB]{252,254,164}\textcolor{black}{0.999} & \cellcolor[RGB]{250,253,160}\textcolor{black}{0.995} & \cellcolor[RGB]{252,254,164}\textcolor{black}{0.999} & \cellcolor[RGB]{249,252,157}\textcolor{black}{0.992} & \cellcolor[RGB]{249,252,157}\textcolor{black}{0.993} & \cellcolor[RGB]{249,252,157}\textcolor{black}{0.994} & \cellcolor[RGB]{247,251,153}\textcolor{black}{0.991} & \cellcolor[RGB]{249,252,157}\textcolor{black}{0.993} & \cellcolor[RGB]{252,254,164}\textcolor{black}{0.999} \\
        rama\_prepro & \cellcolor[RGB]{251,181,26}\textcolor{black}{0.881} & \cellcolor[RGB]{251,157,6}\textcolor{black}{0.844} & \cellcolor[RGB]{245,215,69}\textcolor{black}{0.927} & \cellcolor[RGB]{221,82,56}\textcolor{white}{0.713} & \cellcolor[RGB]{246,213,66}\textcolor{black}{0.925} & \cellcolor[RGB]{234,100,40}\textcolor{black}{0.752} & \cellcolor[RGB]{152,39,101}\textcolor{white}{0.576} & \cellcolor[RGB]{236,103,38}\textcolor{black}{0.757} & \cellcolor[RGB]{243,226,89}\textcolor{black}{0.944} \\
        total & \cellcolor[RGB]{239,109,33}\textcolor{black}{0.768} & \cellcolor[RGB]{245,126,20}\textcolor{black}{0.795} & \cellcolor[RGB]{238,108,34}\textcolor{black}{0.766} & \cellcolor[RGB]{88,16,109}\textcolor{white}{0.464} & \cellcolor[RGB]{99,20,110}\textcolor{white}{0.485} & \cellcolor[RGB]{28,12,67}\textcolor{white}{0.363} & \cellcolor[RGB]{0,0,4}\textcolor{white}{0.285} & \cellcolor[RGB]{8,6,31}\textcolor{white}{0.322} & \cellcolor[RGB]{87,15,109}\textcolor{white}{0.462} \\
        \hline
    \end{tabular}
    }
\end{table}
\begin{table}
    \centering
    \small
    \caption{Pearson correlation coefficients results for all proteins on previously unseen insertion positions ($\text{Test}_{\text{Pos}}$).}
    \label{tab:test2_pearsons}
    \resizebox{\linewidth}{!}{%
    \begin{tabular}{|c|ccc|ccc|ccc|}
        \hline
        \textbf{Rosetta Score} & \multicolumn{3}{c|}{Exhaustive} & \multicolumn{3}{c|}{$\sim$145k Mutations} & \multicolumn{3}{c|}{$\sim$80k Mutations} \\
        & 1crn & 1csp & 1hhp & 2ckx & 1c44 & 5cvz & 1wm2 & 1dc9 & 7dzu \\
        \hline
        fa\_atr & \cellcolor[RGB]{252,254,164}\textcolor{black}{0.998} & \cellcolor[RGB]{252,254,164}\textcolor{black}{0.998} & \cellcolor[RGB]{252,254,164}\textcolor{black}{0.999} & \cellcolor[RGB]{243,226,89}\textcolor{black}{0.947} & \cellcolor[RGB]{243,226,89}\textcolor{black}{0.947} & \cellcolor[RGB]{243,226,89}\textcolor{black}{0.946} & \cellcolor[RGB]{251,185,30}\textcolor{black}{0.889} & \cellcolor[RGB]{243,224,86}\textcolor{black}{0.943} & \cellcolor[RGB]{243,246,137}\textcolor{black}{0.977} \\
        fa\_rep & \cellcolor[RGB]{235,101,39}\textcolor{black}{0.759} & \cellcolor[RGB]{245,126,20}\textcolor{black}{0.800} & \cellcolor[RGB]{228,90,49}\textcolor{white}{0.737} & \cellcolor[RGB]{70,10,105}\textcolor{white}{0.446} & \cellcolor[RGB]{74,11,106}\textcolor{white}{0.453} & \cellcolor[RGB]{20,11,54}\textcolor{white}{0.365} & \cellcolor[RGB]{0,0,3}\textcolor{white}{0.298} & \cellcolor[RGB]{8,6,31}\textcolor{white}{0.338} & \cellcolor[RGB]{15,9,45}\textcolor{white}{0.354} \\
        fa\_sol & \cellcolor[RGB]{252,254,164}\textcolor{black}{0.998} & \cellcolor[RGB]{252,254,164}\textcolor{black}{0.998} & \cellcolor[RGB]{252,254,164}\textcolor{black}{0.999} & \cellcolor[RGB]{242,228,93}\textcolor{black}{0.948} & \cellcolor[RGB]{242,228,93}\textcolor{black}{0.948} & \cellcolor[RGB]{242,228,93}\textcolor{black}{0.949} & \cellcolor[RGB]{251,177,22}\textcolor{black}{0.877} & \cellcolor[RGB]{244,220,79}\textcolor{black}{0.939} & \cellcolor[RGB]{242,244,133}\textcolor{black}{0.977} \\
        fa\_intra\_rep & \cellcolor[RGB]{247,251,153}\textcolor{black}{0.988} & \cellcolor[RGB]{249,252,157}\textcolor{black}{0.994} & \cellcolor[RGB]{250,253,160}\textcolor{black}{0.996} & \cellcolor[RGB]{249,199,47}\textcolor{black}{0.906} & \cellcolor[RGB]{248,135,13}\textcolor{black}{0.814} & \cellcolor[RGB]{251,157,6}\textcolor{black}{0.848} & \cellcolor[RGB]{243,121,24}\textcolor{black}{0.793} & \cellcolor[RGB]{247,131,16}\textcolor{black}{0.810} & \cellcolor[RGB]{250,187,33}\textcolor{black}{0.891} \\
        fa\_intra\_sol & \cellcolor[RGB]{249,252,157}\textcolor{black}{0.993} & \cellcolor[RGB]{252,254,164}\textcolor{black}{0.997} & \cellcolor[RGB]{252,254,164}\textcolor{black}{0.999} & \cellcolor[RGB]{241,237,112}\textcolor{black}{0.963} & \cellcolor[RGB]{241,233,104}\textcolor{black}{0.956} & \cellcolor[RGB]{241,240,121}\textcolor{black}{0.967} & \cellcolor[RGB]{244,219,75}\textcolor{black}{0.935} & \cellcolor[RGB]{241,235,108}\textcolor{black}{0.960} & \cellcolor[RGB]{246,250,149}\textcolor{black}{0.987} \\
        lk\_ball\_wtd & \cellcolor[RGB]{244,247,141}\textcolor{black}{0.983} & \cellcolor[RGB]{242,244,133}\textcolor{black}{0.977} & \cellcolor[RGB]{243,226,89}\textcolor{black}{0.946} & \cellcolor[RGB]{251,183,28}\textcolor{black}{0.885} & \cellcolor[RGB]{250,191,37}\textcolor{black}{0.896} & \cellcolor[RGB]{250,189,35}\textcolor{black}{0.894} & \cellcolor[RGB]{246,128,18}\textcolor{black}{0.804} & \cellcolor[RGB]{248,205,55}\textcolor{black}{0.915} & \cellcolor[RGB]{243,222,82}\textcolor{black}{0.941} \\
        fa\_elec & \cellcolor[RGB]{246,250,149}\textcolor{black}{0.988} & \cellcolor[RGB]{246,250,149}\textcolor{black}{0.987} & \cellcolor[RGB]{249,252,157}\textcolor{black}{0.993} & \cellcolor[RGB]{244,220,79}\textcolor{black}{0.937} & \cellcolor[RGB]{243,222,82}\textcolor{black}{0.941} & \cellcolor[RGB]{247,209,60}\textcolor{black}{0.920} & \cellcolor[RGB]{251,158,7}\textcolor{black}{0.849} & \cellcolor[RGB]{245,217,72}\textcolor{black}{0.932} & \cellcolor[RGB]{242,243,129}\textcolor{black}{0.973} \\
        pro\_close & \cellcolor[RGB]{134,33,106}\textcolor{white}{0.557} & \cellcolor[RGB]{134,33,106}\textcolor{white}{0.555} & \cellcolor[RGB]{187,55,84}\textcolor{white}{0.650} & \cellcolor[RGB]{62,9,102}\textcolor{white}{0.434} & \cellcolor[RGB]{59,9,100}\textcolor{white}{0.429} & \cellcolor[RGB]{15,9,45}\textcolor{white}{0.354} & \cellcolor[RGB]{48,10,92}\textcolor{white}{0.412} & \cellcolor[RGB]{102,21,110}\textcolor{white}{0.502} & \cellcolor[RGB]{83,14,109}\textcolor{white}{0.470} \\
        omega & \cellcolor[RGB]{248,203,52}\textcolor{black}{0.914} & \cellcolor[RGB]{250,149,6}\textcolor{black}{0.837} & \cellcolor[RGB]{250,253,160}\textcolor{black}{0.995} & \cellcolor[RGB]{52,9,95}\textcolor{white}{0.416} & \cellcolor[RGB]{244,220,79}\textcolor{black}{0.937} & \cellcolor[RGB]{141,35,105}\textcolor{white}{0.569} & \cellcolor[RGB]{163,43,97}\textcolor{white}{0.606} & \cellcolor[RGB]{129,31,107}\textcolor{white}{0.549} & \cellcolor[RGB]{246,250,149}\textcolor{black}{0.988} \\
        fa\_dun & \cellcolor[RGB]{249,252,157}\textcolor{black}{0.993} & \cellcolor[RGB]{250,253,160}\textcolor{black}{0.996} & \cellcolor[RGB]{252,254,164}\textcolor{black}{0.998} & \cellcolor[RGB]{241,233,104}\textcolor{black}{0.956} & \cellcolor[RGB]{243,226,89}\textcolor{black}{0.947} & \cellcolor[RGB]{244,220,79}\textcolor{black}{0.938} & \cellcolor[RGB]{242,228,93}\textcolor{black}{0.950} & \cellcolor[RGB]{247,209,60}\textcolor{black}{0.921} & \cellcolor[RGB]{245,248,145}\textcolor{black}{0.985} \\
        p\_aa\_pp & \cellcolor[RGB]{178,50,89}\textcolor{white}{0.635} & \cellcolor[RGB]{249,195,42}\textcolor{black}{0.903} & \cellcolor[RGB]{247,207,58}\textcolor{black}{0.919} & \cellcolor[RGB]{243,121,24}\textcolor{black}{0.792} & \cellcolor[RGB]{247,209,60}\textcolor{black}{0.920} & \cellcolor[RGB]{251,172,16}\textcolor{black}{0.870} & \cellcolor[RGB]{219,79,58}\textcolor{white}{0.717} & \cellcolor[RGB]{248,203,52}\textcolor{black}{0.914} & \cellcolor[RGB]{245,217,72}\textcolor{black}{0.931} \\
        ref & \cellcolor[RGB]{252,254,164}\textcolor{black}{0.999} & \cellcolor[RGB]{250,253,160}\textcolor{black}{0.995} & \cellcolor[RGB]{252,254,164}\textcolor{black}{0.999} & \cellcolor[RGB]{247,251,153}\textcolor{black}{0.991} & \cellcolor[RGB]{249,252,157}\textcolor{black}{0.993} & \cellcolor[RGB]{249,252,157}\textcolor{black}{0.993} & \cellcolor[RGB]{247,251,153}\textcolor{black}{0.990} & \cellcolor[RGB]{249,252,157}\textcolor{black}{0.993} & \cellcolor[RGB]{249,252,157}\textcolor{black}{0.993} \\
        rama\_prepro & \cellcolor[RGB]{250,187,33}\textcolor{black}{0.891} & \cellcolor[RGB]{250,147,6}\textcolor{black}{0.834} & \cellcolor[RGB]{245,215,69}\textcolor{black}{0.929} & \cellcolor[RGB]{224,86,52}\textcolor{white}{0.731} & \cellcolor[RGB]{246,213,66}\textcolor{black}{0.926} & \cellcolor[RGB]{230,94,45}\textcolor{black}{0.747} & \cellcolor[RGB]{147,37,103}\textcolor{white}{0.578} & \cellcolor[RGB]{238,108,34}\textcolor{black}{0.772} & \cellcolor[RGB]{243,224,86}\textcolor{black}{0.942} \\
        total & \cellcolor[RGB]{237,104,37}\textcolor{black}{0.766} & \cellcolor[RGB]{246,129,17}\textcolor{black}{0.805} & \cellcolor[RGB]{236,103,38}\textcolor{black}{0.762} & \cellcolor[RGB]{70,10,105}\textcolor{white}{0.446} & \cellcolor[RGB]{87,15,109}\textcolor{white}{0.476} & \cellcolor[RGB]{22,11,57}\textcolor{white}{0.367} & \cellcolor[RGB]{0,0,4}\textcolor{white}{0.303} & \cellcolor[RGB]{8,6,31}\textcolor{white}{0.337} & \cellcolor[RGB]{20,11,54}\textcolor{white}{0.366} \\    
        \hline
    \end{tabular}
    }
\end{table}

\begin{table}
    \centering
    \small
    \caption{Pearson correlation coefficients results for all proteins on previously unseen amino acid combinations ($\text{Test}_{\text{AA}}$).}
    \label{tab:test3_pearsons}
    \label{}
    \resizebox{\linewidth}{!}{%
    \begin{tabular}{|c|ccc|ccc|ccc|}
        \hline
        \textbf{Rosetta Score} & \multicolumn{3}{c|}{Exhaustive} & \multicolumn{3}{c|}{$\sim$145k Mutations} & \multicolumn{3}{c|}{$\sim$80k Mutations} \\
        & 1crn & 1csp & 1hhp & 2ckx & 1c44 & 5cvz & 1wm2 & 1dc9 & 7dzu \\
        \hline
        fa\_atr & \cellcolor[RGB]{252,254,164}\textcolor{black}{0.997} & \cellcolor[RGB]{252,254,164}\textcolor{black}{0.999} & \cellcolor[RGB]{252,254,164}\textcolor{black}{0.999} & \cellcolor[RGB]{243,224,86}\textcolor{black}{0.946} & \cellcolor[RGB]{243,226,89}\textcolor{black}{0.948} & \cellcolor[RGB]{243,226,89}\textcolor{black}{0.947} & \cellcolor[RGB]{250,187,33}\textcolor{black}{0.895} & \cellcolor[RGB]{244,220,79}\textcolor{black}{0.940} & \cellcolor[RGB]{244,247,141}\textcolor{black}{0.981} \\
        fa\_rep & \cellcolor[RGB]{236,103,38}\textcolor{black}{0.769} & \cellcolor[RGB]{242,116,28}\textcolor{black}{0.790} & \cellcolor[RGB]{226,87,51}\textcolor{white}{0.740} & \cellcolor[RGB]{80,13,108}\textcolor{white}{0.479} & \cellcolor[RGB]{67,10,104}\textcolor{white}{0.456} & \cellcolor[RGB]{13,8,40}\textcolor{white}{0.365} & \cellcolor[RGB]{0,0,3}\textcolor{white}{0.317} & \cellcolor[RGB]{0,0,4}\textcolor{white}{0.321} & \cellcolor[RGB]{64,9,102}\textcolor{white}{0.452} \\
        fa\_sol & \cellcolor[RGB]{252,254,164}\textcolor{black}{0.998} & \cellcolor[RGB]{252,254,164}\textcolor{black}{0.998} & \cellcolor[RGB]{252,254,164}\textcolor{black}{0.999} & \cellcolor[RGB]{243,226,89}\textcolor{black}{0.947} & \cellcolor[RGB]{242,228,93}\textcolor{black}{0.949} & \cellcolor[RGB]{242,228,93}\textcolor{black}{0.949} & \cellcolor[RGB]{251,179,24}\textcolor{black}{0.883} & \cellcolor[RGB]{245,217,72}\textcolor{black}{0.934} & \cellcolor[RGB]{244,247,141}\textcolor{black}{0.982} \\
        fa\_intra\_rep & \cellcolor[RGB]{246,250,149}\textcolor{black}{0.987} & \cellcolor[RGB]{250,253,160}\textcolor{black}{0.994} & \cellcolor[RGB]{250,253,160}\textcolor{black}{0.996} & \cellcolor[RGB]{249,197,44}\textcolor{black}{0.908} & \cellcolor[RGB]{246,128,18}\textcolor{black}{0.810} & \cellcolor[RGB]{251,155,6}\textcolor{black}{0.848} & \cellcolor[RGB]{245,126,20}\textcolor{black}{0.806} & \cellcolor[RGB]{246,129,17}\textcolor{black}{0.811} & \cellcolor[RGB]{248,205,55}\textcolor{black}{0.918} \\
        fa\_intra\_sol & \cellcolor[RGB]{249,252,157}\textcolor{black}{0.994} & \cellcolor[RGB]{252,254,164}\textcolor{black}{0.997} & \cellcolor[RGB]{252,254,164}\textcolor{black}{0.999} & \cellcolor[RGB]{241,235,108}\textcolor{black}{0.962} & \cellcolor[RGB]{241,232,100}\textcolor{black}{0.954} & \cellcolor[RGB]{241,238,116}\textcolor{black}{0.966} & \cellcolor[RGB]{244,219,75}\textcolor{black}{0.936} & \cellcolor[RGB]{241,232,100}\textcolor{black}{0.955} & \cellcolor[RGB]{247,251,153}\textcolor{black}{0.990} \\
        lk\_ball\_wtd & \cellcolor[RGB]{244,247,141}\textcolor{black}{0.983} & \cellcolor[RGB]{242,244,133}\textcolor{black}{0.977} & \cellcolor[RGB]{243,226,89}\textcolor{black}{0.947} & \cellcolor[RGB]{251,176,20}\textcolor{black}{0.879} & \cellcolor[RGB]{250,187,33}\textcolor{black}{0.895} & \cellcolor[RGB]{250,187,33}\textcolor{black}{0.893} & \cellcolor[RGB]{246,128,18}\textcolor{black}{0.808} & \cellcolor[RGB]{249,199,47}\textcolor{black}{0.910} & \cellcolor[RGB]{243,222,82}\textcolor{black}{0.943} \\
        fa\_elec & \cellcolor[RGB]{246,250,149}\textcolor{black}{0.989} & \cellcolor[RGB]{246,250,149}\textcolor{black}{0.987} & \cellcolor[RGB]{249,252,157}\textcolor{black}{0.993} & \cellcolor[RGB]{245,217,72}\textcolor{black}{0.935} & \cellcolor[RGB]{244,220,79}\textcolor{black}{0.940} & \cellcolor[RGB]{248,205,55}\textcolor{black}{0.919} & \cellcolor[RGB]{251,157,6}\textcolor{black}{0.851} & \cellcolor[RGB]{246,211,63}\textcolor{black}{0.927} & \cellcolor[RGB]{242,244,133}\textcolor{black}{0.976} \\
        pro\_close & \cellcolor[RGB]{129,31,107}\textcolor{white}{0.561} & \cellcolor[RGB]{115,26,109}\textcolor{white}{0.536} & \cellcolor[RGB]{175,49,91}\textcolor{white}{0.638} & \cellcolor[RGB]{45,10,88}\textcolor{white}{0.422} & \cellcolor[RGB]{41,11,84}\textcolor{white}{0.418} & \cellcolor[RGB]{9,6,33}\textcolor{white}{0.358} & \cellcolor[RGB]{50,9,93}\textcolor{white}{0.430} & \cellcolor[RGB]{75,12,107}\textcolor{white}{0.471} & \cellcolor[RGB]{90,17,109}\textcolor{white}{0.494} \\
        omega & \cellcolor[RGB]{248,203,52}\textcolor{black}{0.917} & \cellcolor[RGB]{248,138,11}\textcolor{black}{0.826} & \cellcolor[RGB]{250,253,160}\textcolor{black}{0.995} & \cellcolor[RGB]{43,10,86}\textcolor{white}{0.419} & \cellcolor[RGB]{244,219,75}\textcolor{black}{0.937} & \cellcolor[RGB]{126,30,108}\textcolor{white}{0.555} & \cellcolor[RGB]{166,44,95}\textcolor{white}{0.623} & \cellcolor[RGB]{125,29,108}\textcolor{white}{0.552} & \cellcolor[RGB]{249,252,157}\textcolor{black}{0.992} \\
        fa\_dun & \cellcolor[RGB]{249,252,157}\textcolor{black}{0.994} & \cellcolor[RGB]{250,253,160}\textcolor{black}{0.996} & \cellcolor[RGB]{252,254,164}\textcolor{black}{0.998} & \cellcolor[RGB]{241,232,100}\textcolor{black}{0.956} & \cellcolor[RGB]{243,222,82}\textcolor{black}{0.942} & \cellcolor[RGB]{246,213,66}\textcolor{black}{0.930} & \cellcolor[RGB]{244,220,79}\textcolor{black}{0.941} & \cellcolor[RGB]{246,211,63}\textcolor{black}{0.925} & \cellcolor[RGB]{245,248,145}\textcolor{black}{0.984} \\
        p\_aa\_pp & \cellcolor[RGB]{171,46,93}\textcolor{white}{0.631} & \cellcolor[RGB]{249,195,42}\textcolor{black}{0.905} & \cellcolor[RGB]{247,207,58}\textcolor{black}{0.922} & \cellcolor[RGB]{240,112,30}\textcolor{black}{0.785} & \cellcolor[RGB]{247,207,58}\textcolor{black}{0.921} & \cellcolor[RGB]{251,168,13}\textcolor{black}{0.867} & \cellcolor[RGB]{219,79,58}\textcolor{white}{0.724} & \cellcolor[RGB]{247,207,58}\textcolor{black}{0.920} & \cellcolor[RGB]{244,219,75}\textcolor{black}{0.937} \\
        ref & \cellcolor[RGB]{252,254,164}\textcolor{black}{0.999} & \cellcolor[RGB]{250,253,160}\textcolor{black}{0.995} & \cellcolor[RGB]{252,254,164}\textcolor{black}{0.999} & \cellcolor[RGB]{249,252,157}\textcolor{black}{0.993} & \cellcolor[RGB]{249,252,157}\textcolor{black}{0.993} & \cellcolor[RGB]{249,252,157}\textcolor{black}{0.993} & \cellcolor[RGB]{247,251,153}\textcolor{black}{0.991} & \cellcolor[RGB]{249,252,157}\textcolor{black}{0.993} & \cellcolor[RGB]{252,254,164}\textcolor{black}{0.999} \\
        rama\_prepro & \cellcolor[RGB]{251,179,24}\textcolor{black}{0.884} & \cellcolor[RGB]{250,149,6}\textcolor{black}{0.841} & \cellcolor[RGB]{246,211,63}\textcolor{black}{0.927} & \cellcolor[RGB]{217,77,61}\textcolor{white}{0.718} & \cellcolor[RGB]{246,211,63}\textcolor{black}{0.925} & \cellcolor[RGB]{229,91,48}\textcolor{white}{0.747} & \cellcolor[RGB]{144,36,104}\textcolor{white}{0.585} & \cellcolor[RGB]{233,98,42}\textcolor{black}{0.760} & \cellcolor[RGB]{244,220,79}\textcolor{black}{0.940} \\
        total & \cellcolor[RGB]{238,108,34}\textcolor{black}{0.776} & \cellcolor[RGB]{243,119,25}\textcolor{black}{0.795} & \cellcolor[RGB]{234,100,40}\textcolor{black}{0.763} & \cellcolor[RGB]{80,13,108}\textcolor{white}{0.478} & \cellcolor[RGB]{82,14,108}\textcolor{white}{0.482} & \cellcolor[RGB]{14,8,42}\textcolor{white}{0.368} & \cellcolor[RGB]{0,0,4}\textcolor{white}{0.321} & \cellcolor[RGB]{0,0,4}\textcolor{white}{0.320} & \cellcolor[RGB]{72,11,106}\textcolor{white}{0.465} \\        
        \hline
    \end{tabular}
    }
\end{table}
\subsection{Prediction Error Analysis}
\begin{figure*}
    \centering
    \caption{One-to-one plots showing our performance for 1c44 on $\text{Test}_{\text{Rand}}$. The x-axis denotes the true values and the y-axis denotes the predictions made by our PRIMRose model.}
    \includegraphics[width=0.33\textwidth]{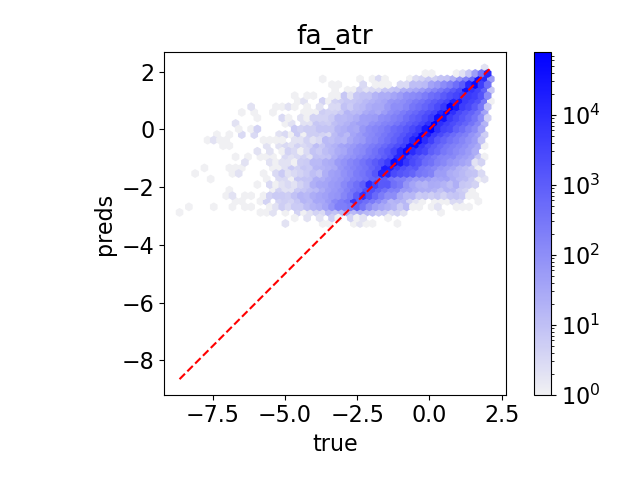}
    \includegraphics[width=0.33\textwidth]{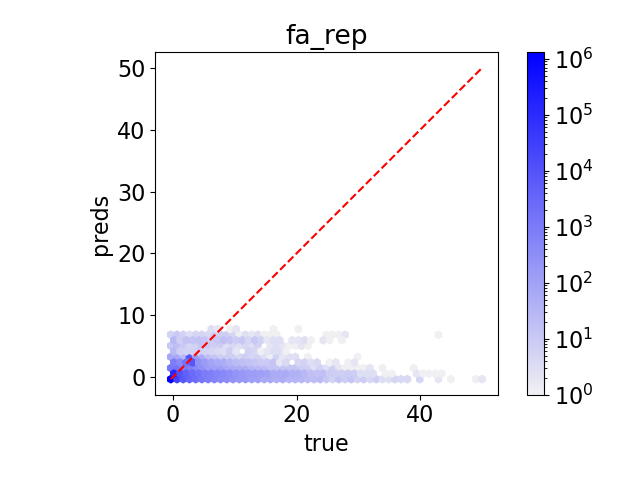}
    \includegraphics[width=0.33\textwidth]{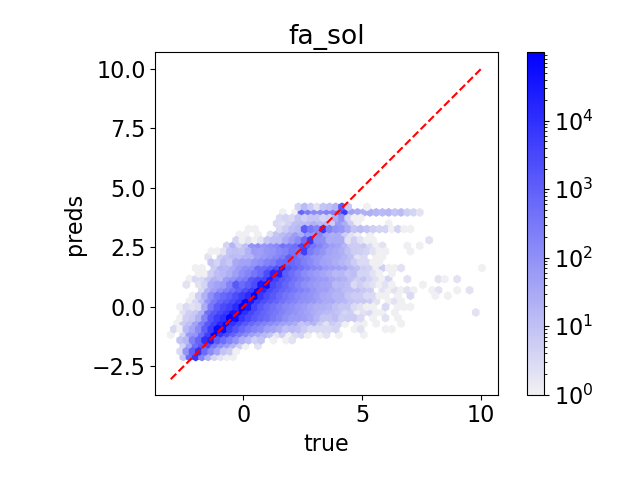}
    \includegraphics[width=0.33\textwidth]{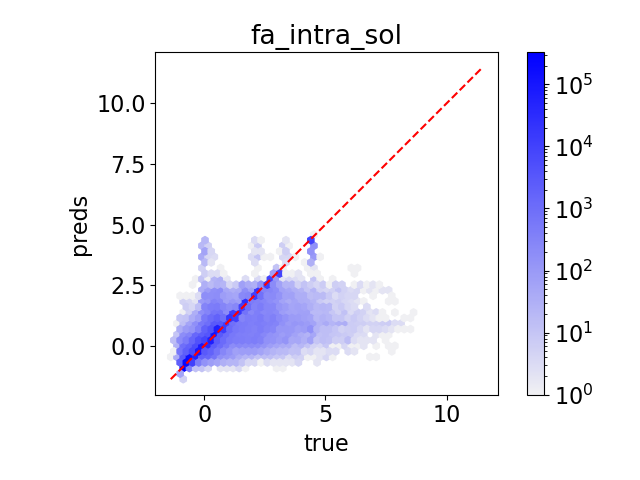}
    \includegraphics[width=0.33\textwidth]{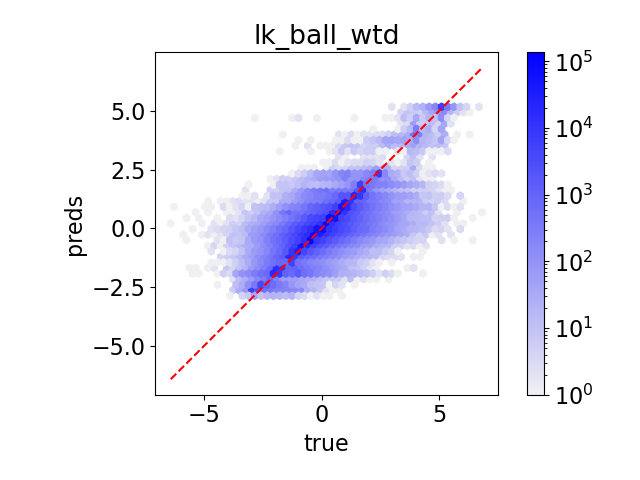}
    \includegraphics[width=0.33\textwidth]{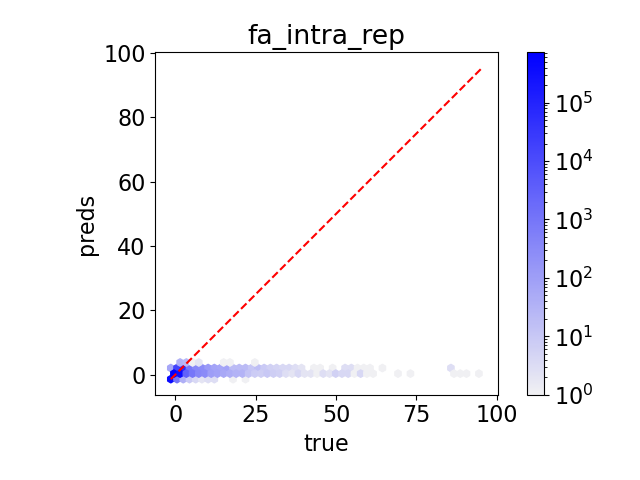}
    \includegraphics[width=0.33\textwidth]{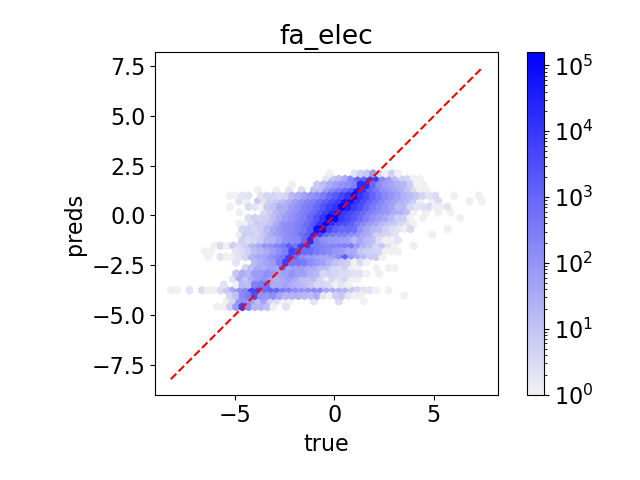}
    \includegraphics[width=0.33\textwidth]{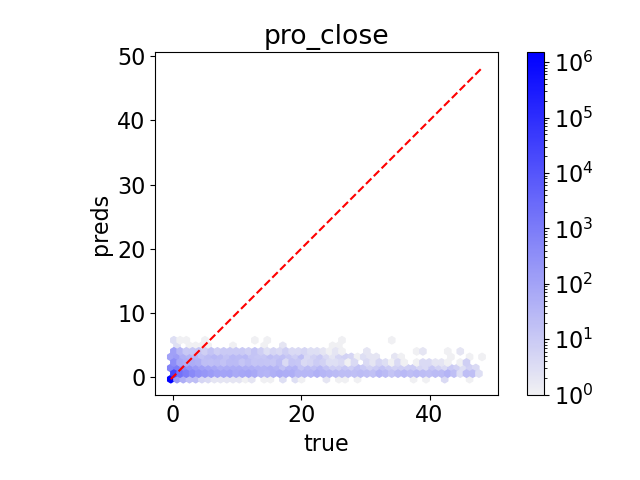}
    \includegraphics[width=0.33\textwidth]{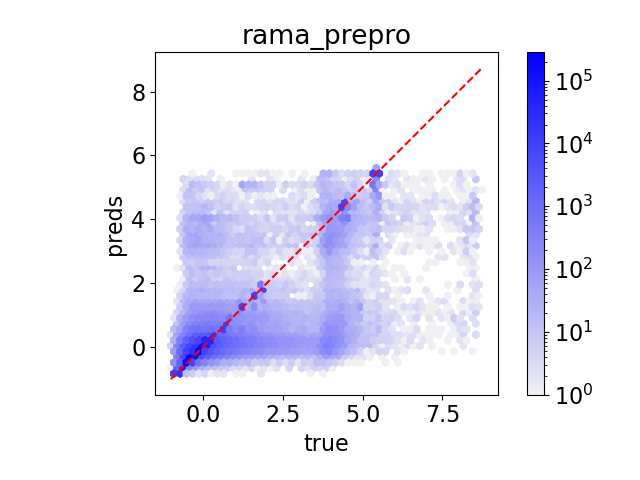}
    \includegraphics[width=0.33\textwidth]{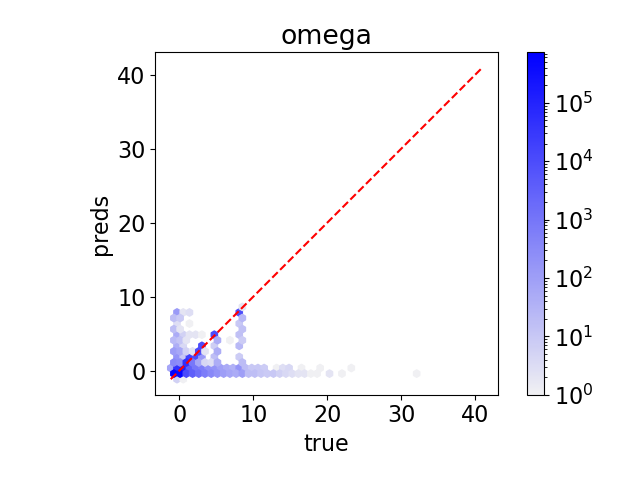}
    \includegraphics[width=0.33\textwidth]{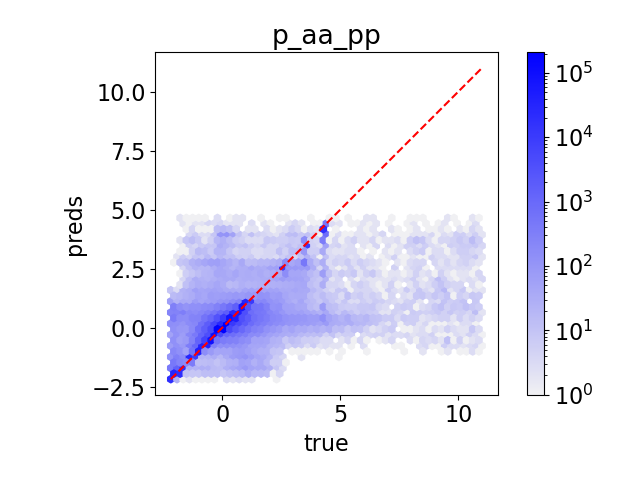}
    \includegraphics[width=0.33\textwidth]{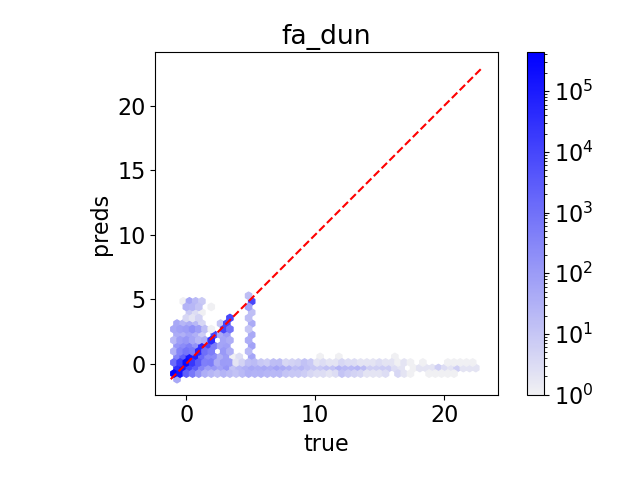}
    \includegraphics[width=0.33\textwidth]{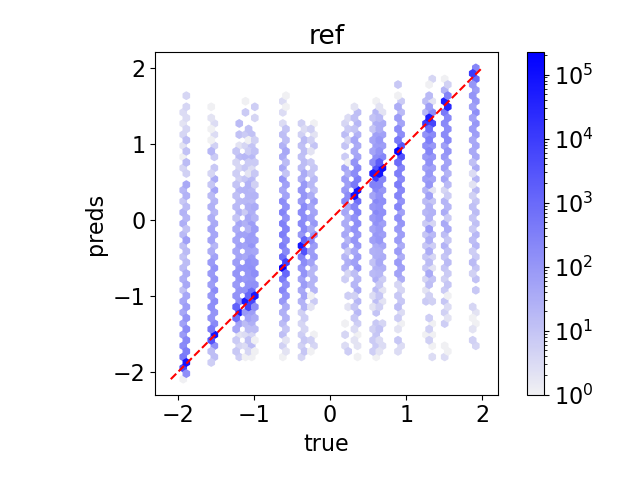}
    \includegraphics[width=0.33\textwidth]{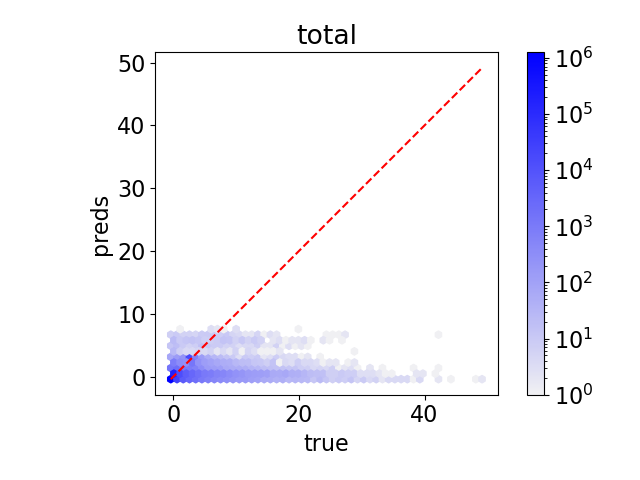}
    \label{fig:1c44_hexbins}
\end{figure*}

In addition to using Pearson correlation coefficients to evaluate model performance, in Fig.~\ref{fig:1c44_hexbins}, we also visualize a one-to-one scatterplot, showing the relationship between predicted values and ground truth values for all $N_pL_p$ values, one score per subfigure. This anslysis is shown only for 
protein 1c44, which exhibited the median $R^2$ across all of our proteins. To better visualize prediction density and identify potential outliers, we utilized a 2D hexagonal binning map with a minimum bin count threshold set to 0.01\% of the dataset. The logarithmic color scale emphasizes the alignment between true values and predicted ref scores.

Notably, the ref plot displays vertical striping, something not seen with the RoseNet model \cite{rosenet_extension}. This striping arises from discretized true labels, while predictions from the ref term remain continuous. Despite this, the overall distribution reflects strong correlation and signal retention in the ref score predictions, supporting the model’s ability to approximate per-residue energy biases across a range of mutation effects.
\subsection{Secondary Structure Analysis}
\begin{figure*}
    \centering
    \caption{Plots showing how our performance varies based on the locations of the insertions. The x-axis denotes the set of all mutants including the given insertion position and the y-axis denotes the average root mean square error over that set. The line style shows the secondary structure at the insertion position.}
    \label{fig:ss_lineplots}
    \includegraphics[width=0.33\textwidth]{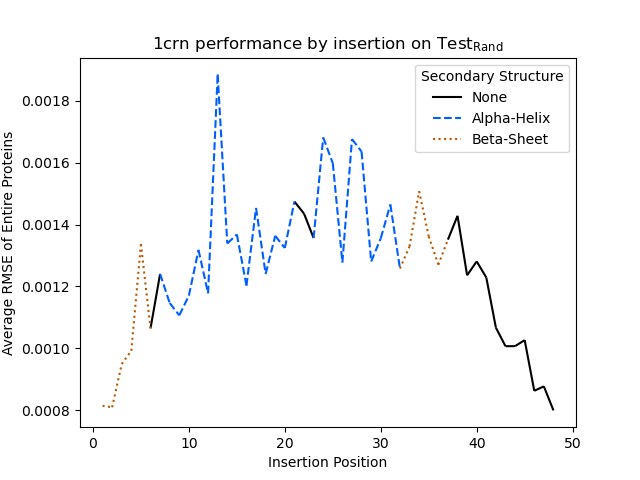}
    \includegraphics[width=0.33\textwidth]{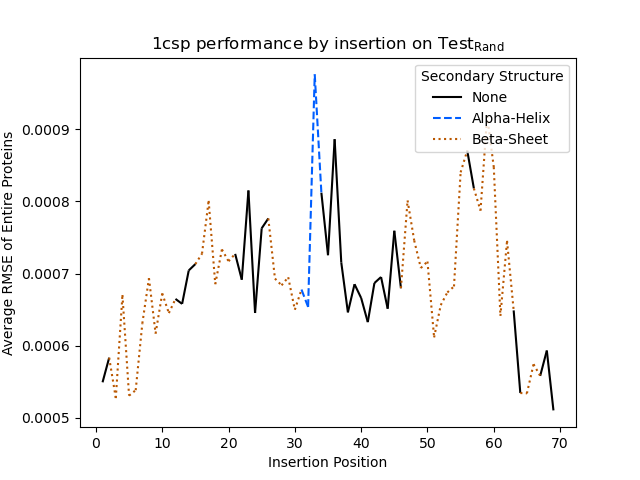}
    \includegraphics[width=0.33\textwidth]{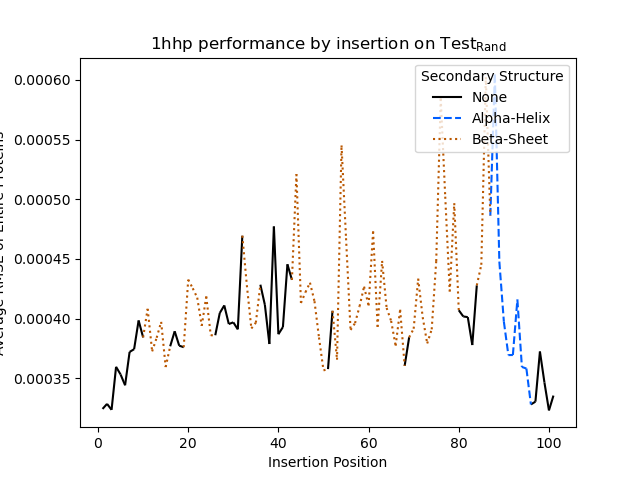}
    \includegraphics[width=0.33\textwidth]{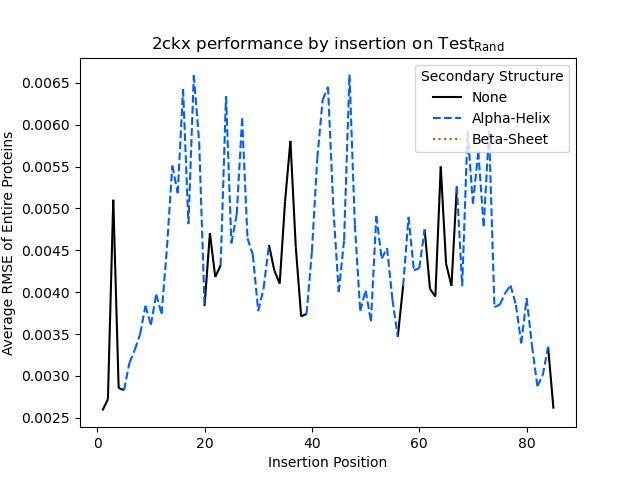}
    \includegraphics[width=0.33\textwidth]{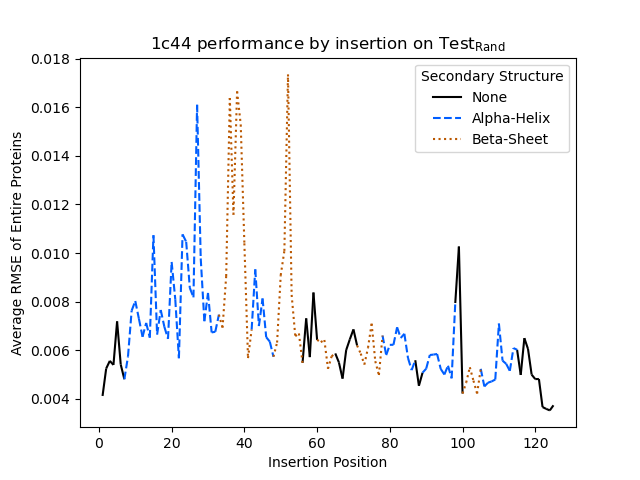}
    \includegraphics[width=0.33\textwidth]{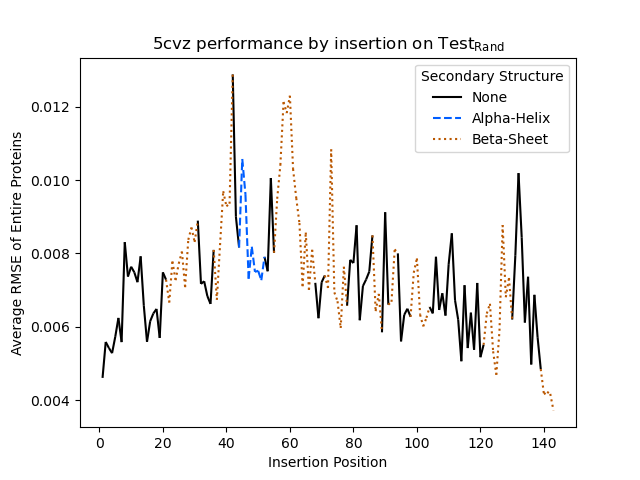}
    \includegraphics[width=0.33\textwidth]{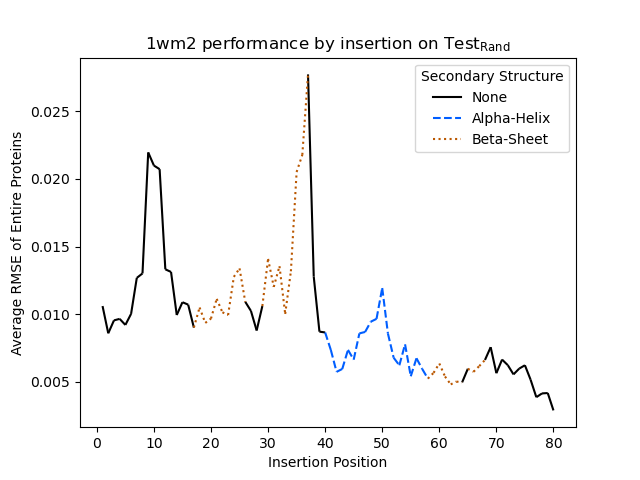}
    \includegraphics[width=0.33\textwidth]{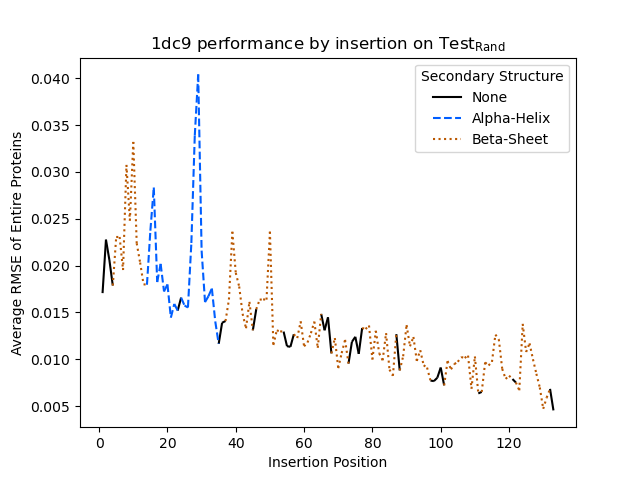}
    \includegraphics[width=0.33\textwidth]{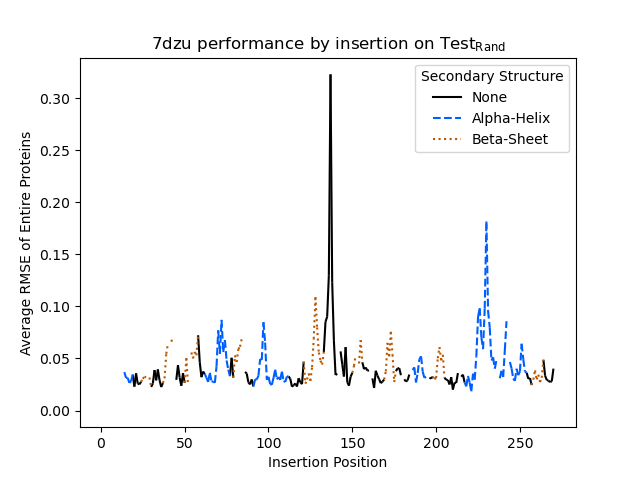}
\end{figure*}

To analyze how the insertion site affects prediction accuracy, we compute the root mean square error (RMSE -- lower is better) across all $L_p$ positions of mutant $m$. In /
Figure~\ref{fig:ss_lineplots}, each line indicates the average RMSE over all mutants that had at least one insertion at that position. That is, a high value at a given position on the line plot indicates relatively poorer predictive performance when an insertion was made at that position.

Across most proteins, insertions at terminal regions (amino-terminus and carboxyl-terminus) result in lower RMSE, indicating that these positions are more predictable.  In contrast, insertions within structured regions (e.g. central alpha-helices, beta-sheets) tend to yield higher error, underscoring the sensitivity of structured cores to energy changes. Some effects vary by protein; in 2ckz, all insertions into its single $\alpha$-helical domain show higher error, while in 1c44, the first alpha-helix appears more critical than others. Other effects, such as inserting into beta-sheets, result in highly unpredictable mutations for PRIMRose across proteins, further revealing the importance of beta-sheets for the overall structure of proteins.

Overall, the effect of inserting into a secondary structure appears to be correlated with the size of the secondary structure as well as the overall size of the protein. These patterns suggest that our model implicitly learns the structural importance and mutational sensitivity associated with secondary structures. The consistency of these trends across proteins demonstrates the model's ability to generalize beyond training mutations and recognize the functional relevance of structural components. 

We should note that we also investigated whether SASA could help explain variation in prediction performance across residues. While SASA is often associated with residue-level stability and mutational sensitivity \cite{Durham2009, Ali2014}, our analysis did not reveal consistent trends across proteins. As such, the results do not offer a meaningful return on interpretative value and are not included here.

\subsection{Computational Cost Analysis}
Generating energy changes with the Rosetta software for protein mutations is a computationally intensive process. Results can take up to weeks for an exhaustive dataset. In contrast, after training for 1 to 6 hours depending on the protein, our model is able to produce scores for mutations in a fraction of the time taken by Rosetta, as shown in Table~\ref{tab:inference_times}.
\begin{table}
    \centering
    \small
    \resizebox{\linewidth}{!}{%
    \begin{tabular}{|c|c|c|c|c|c|c|c|c|c|}
        \hline
        & 1crn & 1csp & 1hhp & 2ckx & 1c44 & 5cvz & 1wm2 & 1dc9 & 7dzu \\
        \hline
        Time (Seconds) & 5.82 & 7.88 & 11.12 & 9.22 & 13.73 & 15.65 & 8.74 & 14.63 & 32.91 \\
  \hline
    \end{tabular}
    }
    \caption{Time required for PRIMRose to produce scores for 50k mutations for each protein with a minibatch of 256.}
    \label{tab:inference_times}
\end{table}

\section{Conclusion and Future Work}
In this work, we introduce PRIMRose, a per-residue deep learning model capable of predicting Rosetta energy values of double InDel mutants with with high accuracy and interpretability. By evaluating across a diverse set of proteins and test conditions - including novel insertion position and amino acid combinations - we demonstrate that our model generalizes well to previously unobserved mutational contexts. Consistently high Pearson correlations across core energy terms reflect the model's strength in capturing local energetic changes, while secondary structure-based analyses reveal its sensitivity to structural context. While the protein's total energy score value shows greater variability across proteins, this reflects the inherent challenges of capturing global energetic effects and emphasizes the benefit of residue-level resolution for understanding mutation effects in detail. These results demonstrate that our model not only accelerates energetic predictions over traditional usage of Rosetta, but also enables biologically interpretable insights into mutational tolerance and highlights structural dependencies within a protein.

There are a number of ways this research could be expanded upon. The existing data sets could be extended by synthesizing mutant data for more proteins using Rosetta. This would allow new models to be trained on any other viable single-chain protein.

Architectural experimentation also has the potential to yield new results. For example, a U-net architecture would consolidate input into shorter sequences for processing before expanding it back to its original size. This would change the way the model propagates information across distances, potentially making it better recognize distant dependencies across input sequences.

Other known residue properties with known or suspected relationships to residue energy metrics or Rosetta scores may be incorporated into the input to improve prediction. The previously mentioned solvent-accessible surface area and secondary structures of each residue, as well as other physicochemical and structural properties, known functional sites and hotspots, may be incorporated into the amino acid embeddings to provide the model with additional information that may produce more accurate or usable results.

While wet lab experiments involving double InDel mutations are relatively rare, comparing PRIMRose predictions with available experimental results — even in limited quantities — can serve as a form of preliminary validation. Additionally, wet lab data on substitution mutations, which are more prevalent in the literature, may offer a useful benchmark for validation or even be leveraged in a pre-training strategy.

Advancements on PRIMRose could extend the model to handle other types of mutations. While our work focuses exclusively on double insertions, relatively simple changes could adapt it to work on double deletions. More complicated problems could include predicting effects of arbitrary combinations of insertions, deletions, and substitutions. Future work may also approach triple or higher-order InDel or arbitrary mutation predictions, although it is important to note that the combinatorial input space of such a problem grows impractically large very quickly, and novel approaches may need to be employed.

Additionally, because PRIMRose takes an entire (variable length) FASTA protein sequence as input, it represents a step towards a more generalizable model that could potentially be trained on multiple proteins. This opens the door for protein-level embeddings that employ rich information extracted from curated bioinformatics databases, including phylogenetic or evolutionary lineage, or gene pathways.

A major benefit of models like PRIMRose is that they can quickly identify promising mutations for future wet lab studies. Future studies may streamline this process by exploring an inverse model that predicts high-impact double InDel mutations. That is, solving this optimization problem:
$$ \arg\max_{x} f(PRIMRose(x)) $$
Where $x$ denotes a FASTA sequence and $f$ is some objective to the be optimized (e.g., maximizing energy change). 
This is a significantly more complicated problem, but would serve to identify high-impact mutations in a single inference pass. 

Finally, to train highly accurate models as efficiently as possible, one could train PRIMRose in an active learning fashion, incrementally running Rosetta to provide the most valuable additional training samples.
\bibliographystyle{ACM-Reference-Format}
\balance
\bibliography{bibliography}

\end{document}